# Significantly increased magnetic anisotropy in Co nano-columnar multilayer structure via a unique sequential oblique-normal deposition approach

Arun Singh Dev[1], Sharanjeet Singh[1], Anup Kumar Bera[1], Pooja Gupta[2,3], Velaga Srihari[4], Pallavi Pandit[5], Matthias Schwartzkopf[5], Stephan V. Roth[5,6] and Dileep Kumar[1*]

[1]**UGC-DAE Consortium for Scientific Research, Khandwa Road, Indore-452017, India**

[2]**Raja Ramanna Centre for Advanced Technology, Indore 452013, India**

[3]**Homi Bhabha National Institute, Training School Complex, Anushakti Nagar, Mumbai 400094, India**

[4]**High Pressure & Synchrotron Radiation Physics Division, Bhabha Atomic Research Centre, Trombay, Mumbai 400085, India**

[5]**Photon Science, DESY, Notkestraße 85, 22607 Hamburg, Germany**

[6]**KTH Royal Institute of Technology, Department of Fibre and Polymer Technology, Teknikringen 56-58, 100 44 Stockholm, Sweden**

*E-mail –dkumar@csr.res.in

Wireless communication devices, high-density magnetic memories, magnetic sensors, and, more recently, spintronic devices are being used and are eager to be used in daily life. Therefore, artificial manipulation of magnetic properties is a prerequisite to increase the quality and performance of such magnetic devices to make them more suitable and easily usable for such applications. Here, a unique oblique-normal sequential deposition technique is used to create Co based multilayer structure [$Co_{oblique}$(4.4nm)/$Co_{normal}$(4.2 nm)]$_{10}$, where each $Co_{oblique}$ layer is deposited at an oblique angle of 75°, to induce large in-plane uniaxial magnetic anisotropy (UMA). Compared to the previous ripple, stress and oblique angle deposition (OAD) related studies on Cobalt in literature, one-order higher UMA with the easy axis of magnetization along the projection of the tilted nano-columns in the multilayer plane is observed. The multilayer retains magnetic anisotropy even after annealing at 450 °C. The in-plane UMA in this multilayer is found to be the combination of shape, and magneto-crystalline anisotropy (MCA) confirmed by the temperature-dependent grazing incidence small angle X-ray scattering (GISAXS), *in situ* reflection high energy electron diffraction (RHEED) and grazing incidence X-ray diffraction (GIXRD) measurements. The crystalline texturing of hcp Co in the multilayer minimizes spin-orbit coupling energy along the column direction, which couples with the shape anisotropy energies and results in preferential orientation of the easy magnetic axis along the projection of the columns in the multilayer plane. Reduction in UMA after annealing is attributed to diffusion/merging of columns and annihilating crystallographic texturing. The obtained one-order high UMA demonstrates the potential application of the unique structure engineering technique, which may have far-reaching advantages in magnetic thin films/multilayers and spintronic devices.

# INTRODUCTION

Magnetic properties, namely uniaxial magnetic anisotropy (UMA) and magnetization reversal process, are the key properties having applications in high-density magnetic memories [1], magnetic sensors [2], spintronic and wireless communication devices [3] etc. As a result, researchers have been developing and exploring various strategies for flexible and fine tailoring of UMA, such as magnetic field and stress annealing [4], external stress [5], interface engineering [6], oblique angle deposition (OAD) [7] and so on. Every technique has its limitations in tailoring magnetic anisotropy. Higher stress, for example, results in buckling, peeling off or cracks in films [8,9,10,11]. The limited penetration depth of ions in the ion beam erosion (IBE) process induces the modifications (ripples) only to the top film surface [12], resulting in limited magnetic anisotropy. Recently we have developed a unique sequential deposition-erosion technique and obtained one-order high UMA in Co-based ripple patterns compared to other ripple-based magnetic studies in the literature [13].

Meanwhile, OAD controls the morphological development of thin films through "surface shadowing" effects which start from the atomic level [14]. Shadowing controls the nucleation of atoms and the nanostructure evolution of the films by preventing the deposition of atoms or molecules in regions situated just behind initially developed nuclei structures [15,16]. Such a process results in tilted columnar film structures with porosity depending on the geometrical condition, such as deposition angle, thickness etc. Structural engineering is not limited only to the tilted nano-columns but also the zig-zag, spiral and s-shaped nano-columns, etc., that can be formed using OAD techniques [14]. Such OAD studies are highly explored in the field of electronics, photonics, biosensors, solar cells, photovoltaic applications, fuel cells, hydrogen storage, Li-ion batteries, piezoelectric nanogenerators, sensors like solid, liquid and gas sensors, pressure sensors and actuators; optical sensors, plasmonics, wetting and micro fluids. All of these applications are very well summarized in an outstanding review article by A. Barranco et al. [14]. Still, the field of OAD is not explored well enough in the area of magnetic thin films.

The role of stress in OAD-based UMA studies is not understood enough in the literature though it creates shape anisotropy. But, in our recent study on OAD FeCo multilayer [17], we have provided the direct role of stress in inducing UMA. However, this study was conducted for a particular oblique incidence of 60°. It is important to note that UMA is not significantly high at lower oblique deposition angles (below 45° OAD) than at higher deposition angles [7,18,19]. Further, no magneto-crystalline anisotropy (MCA) was observed. Despite extensive work on various magnetic polycrystalline films in the literature, the Co-based system remains a model system for studying UMA because Co has excellent growth properties as a thin film, and very flat interfaces can be prepared using sputtering and molecular beam epitaxy techniques [20].

Also, in the case of Co, obliquely deposited films show crystallographic texturing of the (002) peak (c-axis) of the Co hcp structure aligned along the columns direction, which is the easy axis of magnetization hence creating MCA [21,22]. This motivates us to explore the effects of stress, shape and crystallographic texturing on the UMA in Co-based systems at higher oblique deposition angles. Furthermore, it is challenging to keep an easy axis of the magnetic anisotropy in the film/multilayer plane, which is a major concern in Co-based OAD films/multilayers [18]. For example, Co/Cu deposited obliquely at 84° [23,24] and Co/Au deposited obliquely at 45° [25] where the easy axis of magnetic anisotropy is not in-plane or exactly out-of-plane. Despite these big issues, our main motive is to go beyond the general structures formed through OAD to engineer a unique structure that will provide us with significantly increased UMA. Considering all these concerns, it has been planned to explore a Co-based polycrystalline multilayer structure to induce huge UMA and to understand how the structural engineering, artificially created morphology, stress, and crystallographic texturing affect the magnetic behaviour of the multilayer.

Within this context, we have applied a unique approach to deposit Co multilayer, which we have recently developed for FeCo multilayer [17]. It is the oblique-normal sequential depositions to make a multilayer structure. But, in the present study, we increased the deposition angle to 75° OAD and changed the thicknesses slightly. A very strong in-plane UMA is observed in the in-plane projection of vapour deposition direction. The presence of the Bragg peak in the X-ray reflectivity (XRR) pattern confirms the ability of the system to maintain the same morphology up to 10 bilayers, although the system has a complex structure (oblique-normal multilayer structure). The effect of annealing on structure, in-plane UMA, roughness and density of the multilayer has been studied using *in situ* Reflection high energy electron diffraction (RHEED), *ex situ* magneto-optical Kerr effect (MOKE), and XRR, respectively. Furthermore, atomic force microscopy (AFM) as well as synchrotron-based grazing incidence X-ray diffraction (GIXRD) and grazing incidence small angle X-ray scattering (GISAXS) measurements are performed to study the temperature-dependent structural and morphological evolution in the multilayer and pretty well correlated with the observed huge UMA.

## EXPERIMENTAL DETAILS

The Co multilayer structure [Co 4.4 nm; OAD=75° /Co 4.2 nm; NAD]$_{10}$ is deposited on SiO$_2$/Si(100) substrates under base pressure of $2 \times 10^{-7}$ mbar using e-beam evaporation as a deposition source. NAD represents the normal angle deposition (normal deposition). Figure 1(a) shows the schematic of deposition geometry, while the final multilayer is shown in fig. 1(b). 75° and 0° angles of incident flux with respect to the substrate normal are used alternatively as the deposition angles for the incident evaporated material to get a multilayer structure. The thickness of each layer is monitored using calibrated quartz crystal monitor in both configurations: OAD and NAD. Structural information using RHEED patterns is collected *in situ* in a UHV chamber [26] at a base pressure of $4\times10^{-10}$ mbar while annealing the Co multilayer to 500 °C. The magnetic characterization of pristine and annealed samples is performed using *ex situ* MOKE measurements. These samples are studied using Bruker D8 diffractometer having Cu Kα radiation for *ex situ* XRR measurements to obtain the structural information: electron density (ρ) and roughness (σ) of the multilayer. All these measurements make it possible to correlate the UMA with the multilayer structure and morphology with increasing temperature.

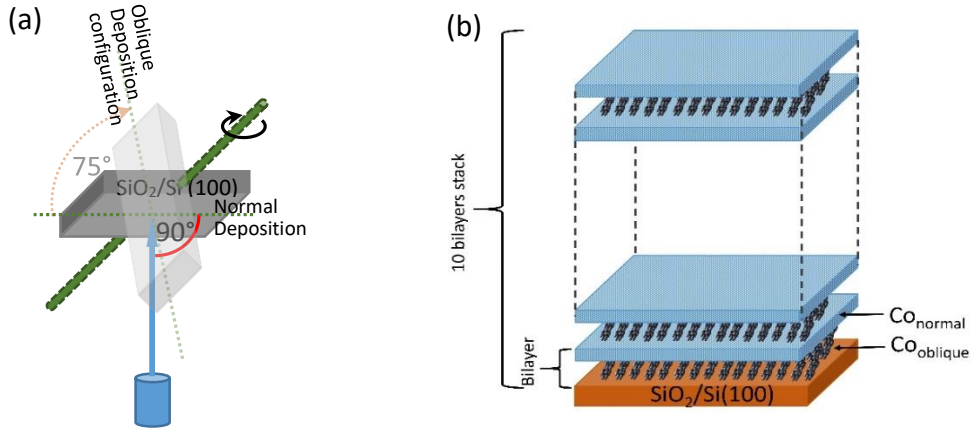

**Figure 1:** (a) Schematic diagram of the deposition geometry for both normal and oblique deposition of Co layers, green rod shows the axis about which the substrate is rotated (b) schematic representation of final multilayer having 10 bilayers stack (Co multilayer) on Si (100) substrate containing native oxide layer. A tilted atomic columnar structure represents an oblique layer.

GISAXS measurements are performed at the P03 (MiNaXS) beamline, Petra-III storage ring at DESY, Hamburg (Germany). An incident angle of 0.45° and a photon energy of 13 keV (λ=0.095373 Å) is used for such measurements. The platform of portable mini X-ray diffraction (XRD) chamber is used for GISAXS measurements to observe the morphology of the samples already annealed *in situ* at temperatures up to 500 °C. The chamber's complete detail and working geometry are already published [27]. PILATUS 1-M (Dectris Ltd., Switzerland) detector, which has a pixel size of $(172 \times 172)$ μm$^2$, is used for GISAXS data recording. AFM studies are also carried out separately to observe morphology. GIXRD measurements of the pristine and annealed samples are done with an X-ray beam of wavelengths λ=0.505917 Å and λ=0.78445 Å at X-ray synchrotron BL11, Indus2, RRCAT (India). Scattered X-ray intensity is mapped for grazing incident angle of ~0.4° using a 2D image plate detector to receive information about the microstructure along the OAD column direction.

## RESULTS

MOKE loops are taken in the longitudinal geometry, and results are shown in fig. 2(a) to 2(d). At room temperature (pristine sample), when the field is applied along the direction of the in-plane projection of the columns in the multilayer plane, the sheared hysteresis loop is obtained, which is the easy axis. At the same time, it becomes rounded, i.e. hard axis, when the field is applied along the perpendicular direction. The magnetic field is insufficient to saturate the moments in any direction for the pristine sample (fig. 2a).

The hysteresis loop along the in-plane projection of the columns becomes a perfect square for the 350 °C annealed sample (fig. 2b). The steepness of hard loops gradually increases with increasing temperature, and at 500 °C (fig. 2d), it transforms to near square loops, which means UMA diminishes with annealing. Coercivity continuously decreases up to 450 °C (fig. 2a to 2c) and then increases at 500 °C (fig. 2d), as plotted in fig. 2(e). Magnetic anisotropies are calculated using the area difference method (difference in areas of easy vs hard loops) [5] and plotted vs temperature in fig. 2(e) and shown in table 1, which confirms the significant reduction in anisotropy at 500 °C.

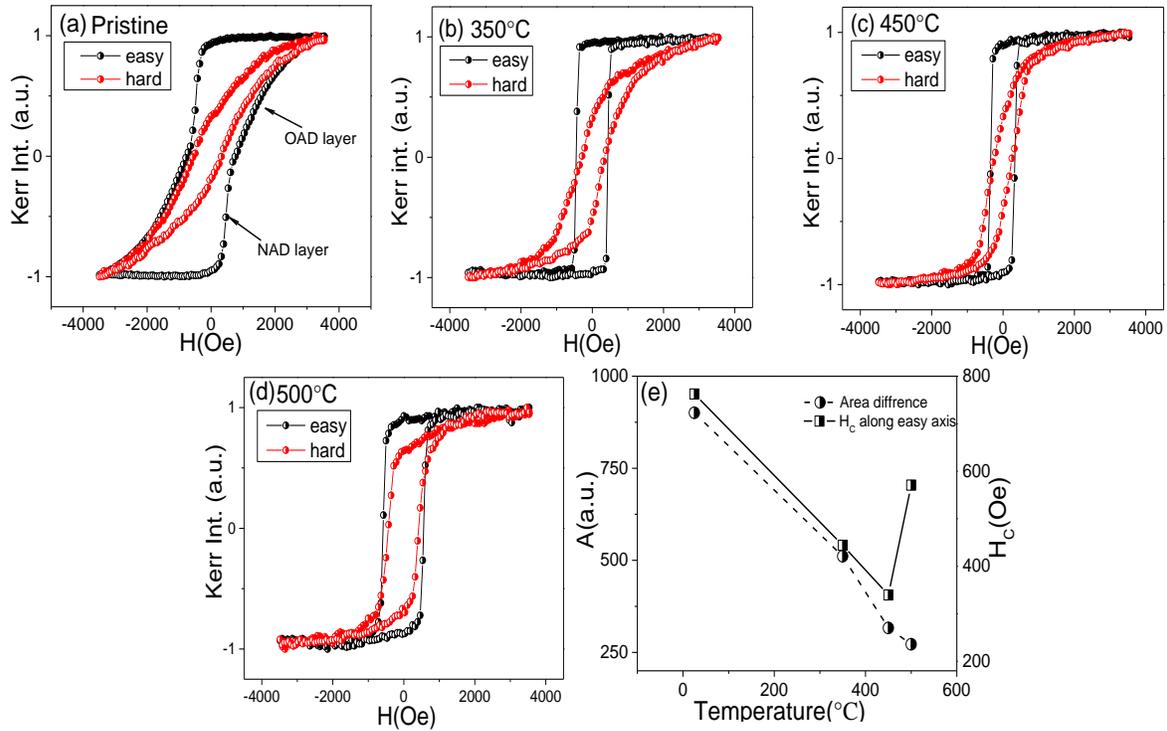

**Figure 2:** Hysteresis loops for (a) pristine and annealed samples at (b) 350 °C, (c) 450 °C and (d) 500 °C, along the easy and hard axes while (e) shows the area difference (A) between the easy and hard axis and $H_C$ along the easy axis at various temperatures. In (a), parts of the loop shape caused by NAD and OAD layers are marked.

| Temperature | Area difference A (a.u.) |
|---|---|
| Room temperature | 900.60 |
| 350 °C | 511.03 |
| 450 °C | 316.29 |
| 500 °C | 272.09 |

**Table 1:** Area difference A vs temperature.

To obtain crystallographic information, *in situ* RHEED measurements are performed by projecting e-beam along θ=0°, and θ=90° to the in-plane projection of columns as shown in fig. 3. Main features of the RHEED measurements are as follows: (i) Concentric rings confirm that structure grows in the polycrystalline hcp phase, (ii) Texturing is visible when columns direction is placed perpendicular to the e-beam direction, (iii) Texturing in the hcp (002) peak is higher along the direction marked by the blue arrow. This arrow makes an average angle φ~40° with respect to the substrate's normal, (iv) at a temperature of 500 °C (fig. 3f), texturing disappeared. In the case of Co OAD, the direction of the hcp c-axis is found to be along the column direction, which generally contributes to the easy axis of the MCA [21,22,28].

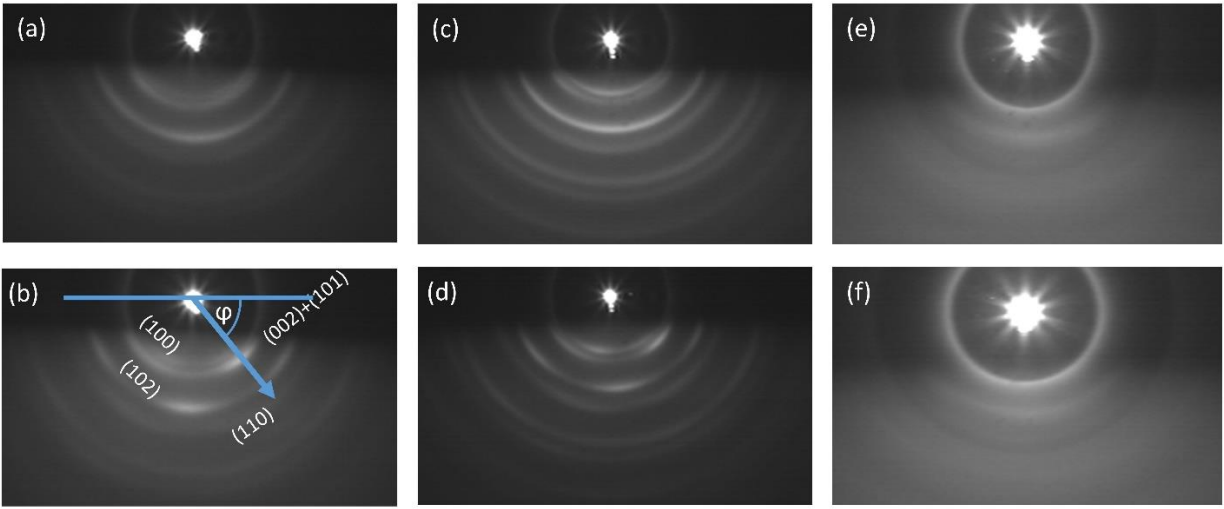

**Figure 3:** RHEED images along (a),(c),(e) θ=0° and (b),(d),(f) θ=90° at (a),(b) room temperature, (c),(d) 400 °C and (e),(f) 500 °C.

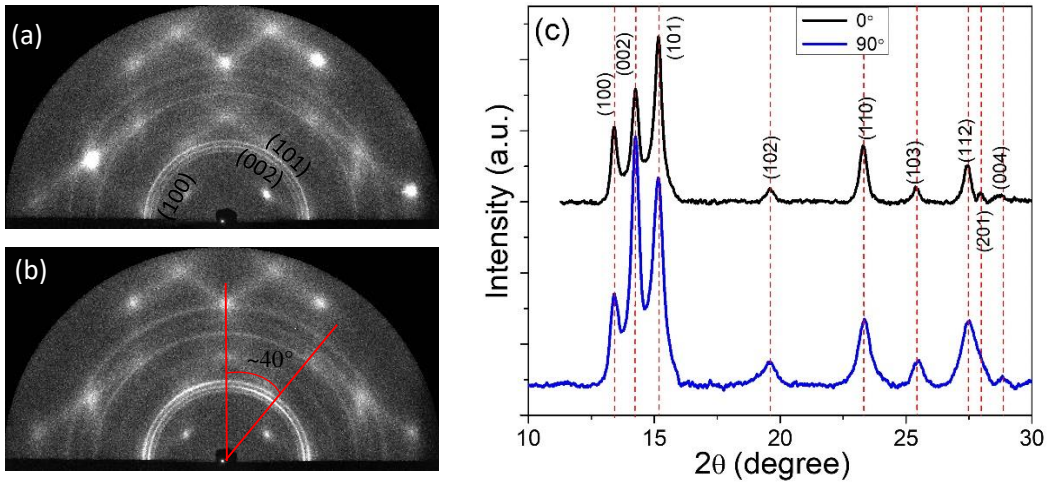

**Figure 4:** GIXRD image (for X-ray λ=0.505917 Å) of pristine Co multilayer along (a) θ=0° and (b) θ=90° directions (c) corresponding 1D plots.

RHEED is a surface-sensitive technique, so depth-resolved structure analysis is performed. Therefore, *ex situ* GIXRD measurement for the pristine sample is done along two perpendicular directions similar to RHEED and shown in fig. 4(a) and 4(b). X-ray wavelength λ=0.505917 Å is used for the measurements. Concentric rings correspond to hcp Co, while bright spots correspond to the Si substrate. Similar weak texturing (as the ring is continuous) is found for the hcp (002) peak for GIXRD taken along θ=90° direction (fig. 4b), confirming that weak texturing is present throughout the depth in the OAD layer. The average texturing angle is similar to RHEED measurements (φ~40°). The GIXRD 1D plots are shown in fig. 4(c), and we draw two conclusions from it: i) along θ=90°, the relative peak intensity of the hcp (002) peak is higher as compared to the GIXRD along θ=0°. It means the hcp (002) peak is textured for GIXRD measurements taken along θ=90° direction. ii) No peak shift found along both directions confirms the absence of stress in the multilayer.

Furthermore, we just rotated the θ=90° position of the sample by 180°. In this position, columns again remain perpendicular (θ=90°) to the X-ray beam direction, but the direction of texturing should turn opposite in the GIXRD image. We performed this measurement at an X-ray wavelength λ=0.78445 Å. The measurements are carried out for pristine and annealed samples (400 °C, 500 °C) along θ=0° and θ=90° to the in-plane projection of columns, as shown in fig. 5(a) to 5(f). As expected, texturing is observed opposite for θ=90° positions (fig. 5b and 5d). At 400 °C, additional in-plane texturing is observed for θ=90° [marked by blue arrows (near in-plane) in fig. 5(d)]. The extra peaks are observed at a temperature of 400 °C and above. The 1D data (fig. 5g) is extracted from the measurements along the θ=0° direction, and it is found that the fcc phase starts to grow at 400 °C and becomes dominant at 500 °C as the peak heights of the hcp phase are significantly decreased at 500 °C. Both hcp and fcc phases coexist up to 500 °C. Structural change considerably reduces texturing in the hcp (002) peak at 500 °C (fig. 5f). Crystallite size is also calculated from the fitting of GIXRD 1D data at various temperatures. For the hcp (002) peak, crystallite size at room temperature, 400 °C and 500 °C is found to be 5.4 nm, 8.3 nm and 15.2 nm, respectively.

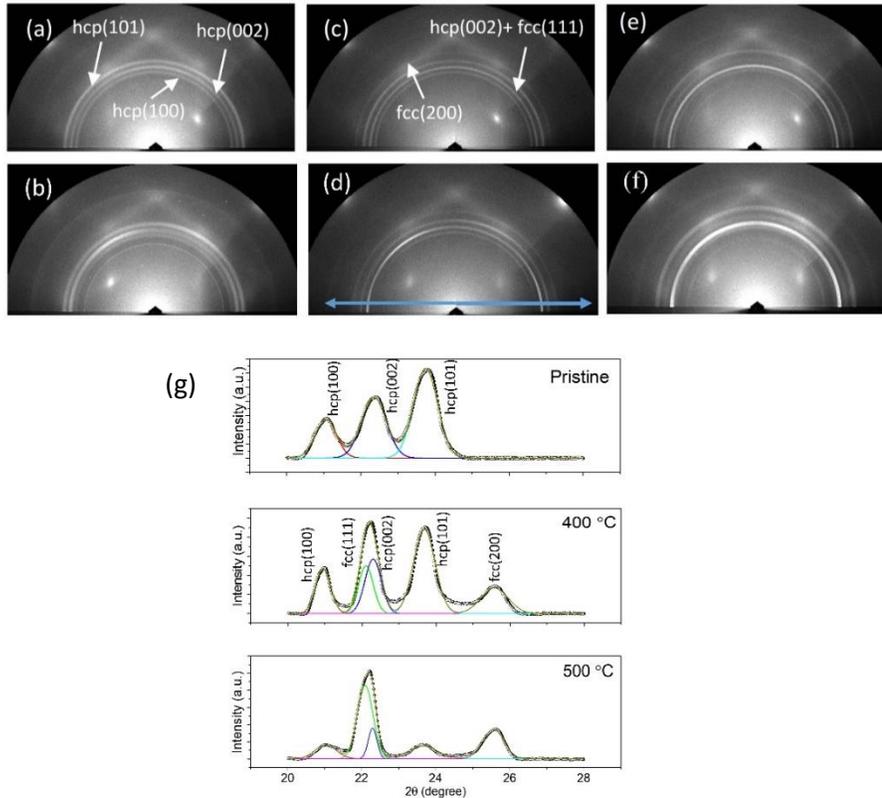

**Figure 5:** Temperature-dependent GIXRD (for X-ray λ=0.78445 Å) images at (a-b) room temperature and for annealed samples at (c),(d) 400 °C and (e),(f) 500 °C for X-ray along (a),(c),(e) θ=0° and (b),(d),(f) θ=90° to the columns (in-plane projection), (g) fitting of the 1D data shows phase change.

XRR measurements are done for the pristine and the UHV annealed sample at 500 °C. The outcome of the performed XRR measurement is shown in fig. 6(a). A clear Bragg peak can be seen for the pristine sample, confirming the multilayer structure of the sample with layers of different electron densities despite the fact that only Co was used for deposition. The difference in the densities hence multilayer formation is the outcome of the deposition method we used: alternate OAD and NAD. We know that the OAD forms the layer having oblique or tilted nano-columns, which results from the shadowing effect [14] and hence possesses the porous layer with lower electron densities. In contrast, the NAD layer possesses a higher electron density.

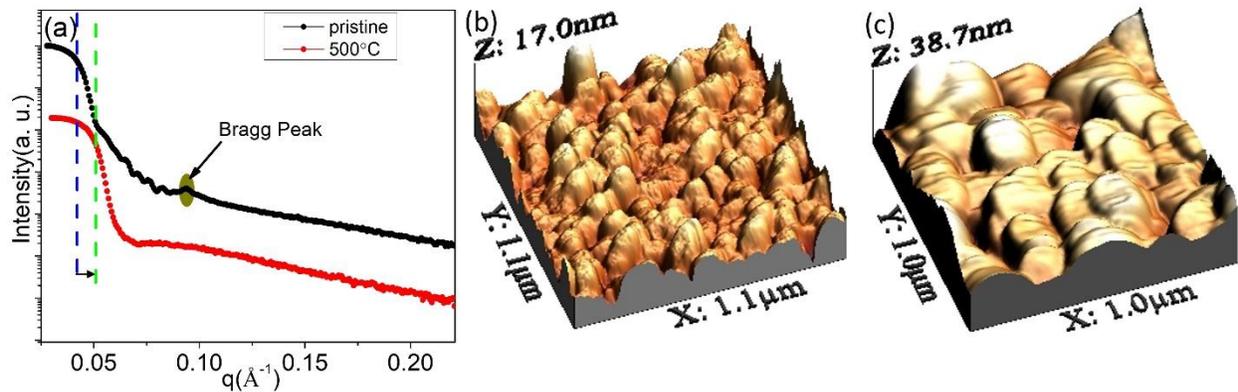

**Figure 6:** (a) XRR plots of pristine and 500 °C annealed samples of Co multilayer. AFM images of (b) pristine and (c) 500 °C annealed samples. The x-axis represents the in-plane projection of the OAD (θ=0°).

The annealed multilayer doesn't show any clear thickness oscillations. The possible reason for such oscillation-less XRR may be the increased roughness after annealing. Also, it can be seen from the XRR plots that the overall electron density of the multilayer is increased (increased critical angle: shifting from dashed blue to the green line in fig. 6a) and reaches near the bulk Co density, which suggests the diffusion of the OAD layer. This process causes the diminishing of the Bragg peak and turns the multilayer into a single-layer film. Another reason for the absence of the Bragg peak can be the increased roughness as XRR of the annealed sample shows no oscillation, but increased electron density is a clear sign of Bragg peak disappearance at 500 ºC.

The 3D AFM images of the pristine and 500 ºC annealed sample are presented in fig. 6(b) and 6(c) respectivelly. The grains are marginally elongated along the in-plane direction of the OAD ($\theta=0°$), but the tilted columnar structure can't be seen. It may be due to the reason that the top layer is normally deposited. The annealed sample shows a larger grain size without elongation in any direction. The rms roughness for the pristine and annealed samples is found to be 2.4 nm and 7.6 nm, respectively, using WSXM software [29] (2D AFM images with roughness plots are included in supplementary material). This increased roughness matches the XRR plot for annealed sample, as it doesn't show any clear thickness oscillations.

For more insights into the buried nanostructures and their temperature-dependent morphological structure variation, GISAXS measurements are performed separately. Figure 7(a) represents the geometry of the measurements using an X-ray beam of wavelength $\lambda=0.95373$ Å with incidence angle $\alpha_i=0.45º$ from the sample surface. $k_i$ and $k_f$ are the incident and reflected wave vectors. The vertical pink plane of tilted columns is kept perpendicular (i.e. $\theta=90°$, similar to RHEED and GIXRD) to the plane containing vectors $k_i$ and $k_f$. In GISAXS measurement, the information regarding vertical and lateral film structure is obtained by the vertical cut along $q_z$ at a non-specular intensity and horizontal cut taken at the Yoneda peak region along $q_y$ direction, respectively.

The 2D GISAXS images of samples annealed at various temperatures are shown in fig. 7(b) to 7(f). The stopper for Direct beam and specular reflection at $q_y=0$ nm$^{-1}$ and along $q_z$ direction are used to avoid saturation of the GISAXS detector. As we explore the obtained GISAXS results, The GISAXS image for the pristine sample (fig. 7b) shows clear asymmetry in scattered intensity distribution about the specular rod ($q_y=0.0$ Å$^{-1}$). The asymmetric distribution of scattered intensity remains present for the multilayer annealed up to 450 ºC (fig. 7e) and almost vanished for the sample annealed at 500 ºC (fig. 7f).

Two crossed elongated intensities are observed for the pristine sample: (i) elongated intensity I, marked by a dashed red line (fig. 7b), is the reflection of the X-ray beam from face 1 of the columns (fig. 7g) where the angle of the dashed red line made with horizontal direction provides the angular tilt ($\beta$) of the columns with respect to the substrate normal and, this type of elongated intensity is also visible on direction beam position (marked by the dashed white line in fig. 7b), (ii) elongated intensity II (dashed blue line in fig. 7b), which are the reflection from the face 2 of the columnar structures (fig. 7g). Additional elongated intensity observed is the elongated intensity III, which comes from the scattering through both NAD and OAD layers. The elongated intensities in the GISAXS data represent the multilayer's steep (or tilted) structures. The images also clearly show that the present asymmetry in intensity almost disappeared for the sample annealed up to 500 °C (fig. 7f). For the pristine sample (fig. 7b), the tilt angle between columns and substrate normal is observed as β~40º, while the deposition angle is α=75°. This inequality (α≠β) arises due to adatom mobility, a highly material-dependent property [30,31].

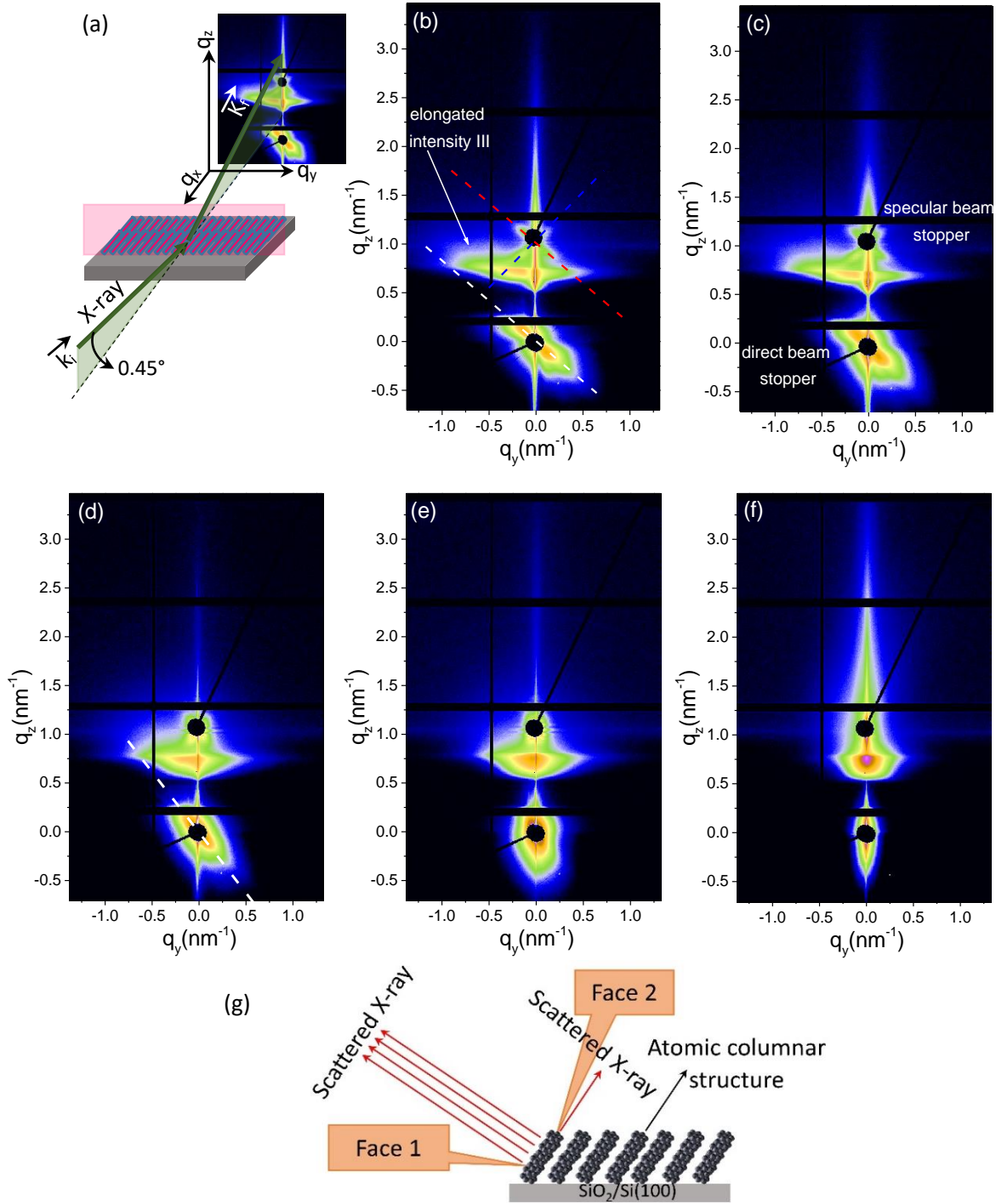

**Figure 7:** (a) GISAXS geometry, 2-D GISAXS images for (b) pristine and annealed samples at (c) 250 °C, (d) 350 °C, (e) 450 °C and (f) 500 °C. (g) Both the faces of the oblique columns scatter the X-ray in respective directions (the direction of the incident X-ray is perpendicular to the plane containing the columns).

1D profiles (I vs q plots) are obtained from the 2D GISAXS patterns to find the in-plain and out-of-plain correlation with increasing temperature, as shown in fig. 8(a) and 8(b). Figure 8(a) clearly shows that the I vs $-q_y$ variation significantly differs from the I vs $+q_y$ variation. Intensity maxima along the $-q_y$ direction are visible. This intensity maxima (shoulder-like satellite peak) is relatively intense compared to the maxima peak at the right, which is hardly visible. The intensity maximum represents the internal columnar or lateral distance between the columnar structures. The intensity maximum at $q_y \approx -0.25$ nm$^{-1}$ for the pristine sample corresponds to the lateral spacing of $\approx 25.1$ nm. As the temperature increases, intensity maxima shift toward lower $q_y$ values (i.e. towards Yoneda peak) for temperature above 250 °C (peak shifts from vertical black line to the blue line in fig. 8a) and become weaker when temperature increases to 500 °C. The peak shift confirms the increasing separation (centre to centre) of the columns confirming that columns are diffusing into the bigger columns and eventually diffusing to form a column-less layer. It is to be noted that a horizontal cut at the Yoneda peak region (in our case at $q_z=0.75$ nm$^{-1}$) along $q_y$ direction is used to analyze the lateral separations of formed structures in the films [17]. But it is clear to see the corresponding fig. 7(f) for 500 °C at the region near $q_z=1.0$ nm$^{-1}$ and along $q_y$ direction that the intensity distribution is still not fully symmetric, but we can treat it as almost symmetric, meaning the columnar structure is almost destroyed at 500 ºC.

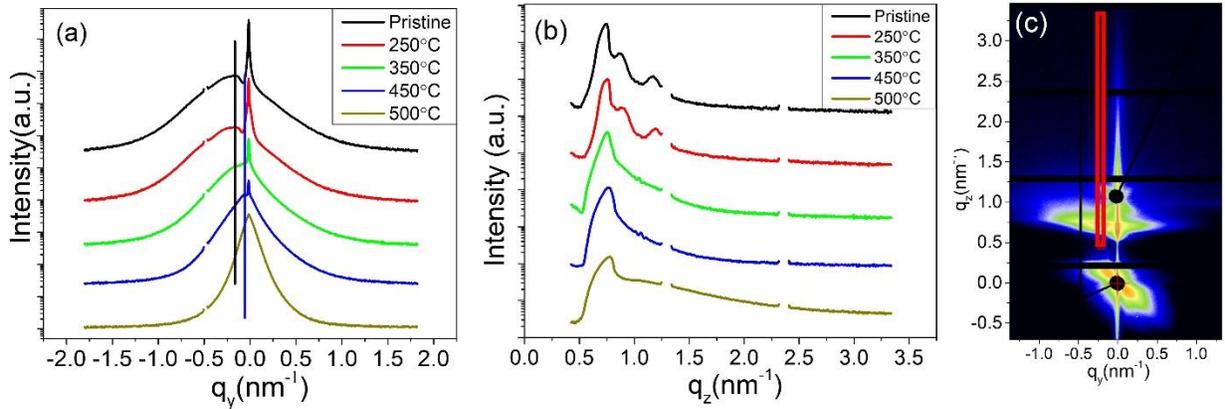

**Figure 8:** (a) horizontal 1D GISAXS line plot for respective temperatures at Yoneda wing region (b) vertical 1D GISAXS line plot for respective temperatures for the respective vertical red line in (c).

To analyze verticle intensity oscillations, the 1D I vs $q_z$ plot (fig. 8b) is derived from verticle cuts (vertical red rectangle in fig. 8c) marked near the satellite peak (off specular position). Though verticle intensity oscillations are not visible in the GISAXS patterns (fig. 7b to 7f) compared to another GISAXS report [32], the interference fringes are visible as damped intensity oscillations in fig. 8(b). The damping of the oscillations is a measure of the surface topography correlation. Slow damping corresponds to the high conformity of the film/multilayer. But in our case, high damping shows no correlation of the multilayer with the substrate because, in the present case, the multilayer is a complex structure containing oblique-normal bilayers, and the Si substrate is not patterned as in the literature [32]. But we can see in fig. 8(b) that with increasing temperature up to 250 ºC, the presence of oscillations symmetry shows that the multilayer still resembles the modulation of pristine multilayer, which matches with the column diffusion information obtained through the explanation of fig. 8(a) that up to the temperature of 250 °C, columns are not diffused.

## DISCUSSION

In oblique angle deposition, in the initial stage of film growth due to the self-shadowing effect, the Volmer–Weber type of growth is expected [33]. In the present case, Co islands nucleate on the Si substrate at various nucleation centres. These islands, with further deposition, transform into elongated columns with different crystallographic orientations due to OAD. In the present Co multilayer, the stress is found to be absent. The stress is observed in some of the OAD studies in the literature [17,34], but the Co-related Co/Cu and Co/Au multilayers [23,24,25,35] don't mention the stress. Possible reasons for the absence of stress in the present multilayer structure are the different deposition angles (75°: higher deposition angle), the material used, deposition rate and thickness etc. [36,37]. In general, OAD Co (hcp) film contains MCA due to texturing of the hcp (002) peak (or [002] plane) along the column direction [21,22,28]. In our study, we also have observed the weak texturing of this hcp (002) peak along the column direction as confirmed by RHEED, GIXRD and GISAXS measurements, as the angle of texturing $\varphi \sim 40°$ is the same as the angle of tilt $\beta \sim 40°$ of columns from the substrate normal. Also, it is a well-established fact that OAD provides shape anisotropy whose strength depends on the angle of deposition [18,19,38,39,40,41,42] and film thickness [39,43]. The oblique deposition above $\alpha=60°$ causes an easy axis of UMA along the in-plane direction of the OAD projection [18,19,39,40,41,42,43]. Therefore, the observed magnetic anisotropy in the present Co multilayer is mainly due to the shape anisotropy generated through OAD and weak texturing in the hcp (002) peak in the multilayer structure. The overall minimization of the weak magneto-crystalline and strong shape anisotropy energy results in the alignment of the corresponding easy axes of magnetization in the in-plane projection of the OAD direction.

This hysteresis loop of such shape along the easy axis for pristine sample (fig. 2a) is also found in literature in Co/Cu multilayer deposited obliquely [23,24], but it does not explain the cause of such a shape. Knepper et al. [44] attribute it to the interlayer coupling in Co/Pt multilayer. Similarly, Jamal et al. [45] refer to such shape as an interlayer coupling in FeCoB/MgO/FeCoB trilayer. Here, in the present case, the moments along the c-axis of hcp $Co_{oblique}$ layers are coupled through the $Co_{normal}$ layer (in-plane moments) and show such behaviour. The hysteresis behaviour observed along the easy axis can be explained by the superposition of two hysteresis loops corresponding to two magnetic components in the OAD and NAD layers, respectively. Contributions of two magnetic components are separated out by fitting loops with a mathematical function [46,47] by taking Kerr signal K(H), saturation magnetization (Ms), coercive field (Hc), squareness (Mr), and n=2 for two magnetic components.

$$K(H) = K(H)_{nor} + K(H)_{obl}$$

$$K(H) = \frac{2M_s^{nor}}{\pi} \arctan \left| \frac{(H \pm H_c^{nor})}{H_c^{nor}} \tan\left(\frac{\pi M_r^{nor}}{2}\right) \right| + \frac{2M_s^{obl}}{\pi} \arctan \left| \frac{(H \pm H_c^{obl})}{H_c^{obl}} \tan\left(\frac{\pi M_r^{obl}}{2}\right) \right| \quad (1)$$

where 'nor' and 'obl' represent normal and oblique layers, respectively. The values of these parameters are taken by assuming a square loop for Co$_{normal}$: $M_r^{nor} = 0.92$, $M_s^{nor} = 0.5$, and $H_c^{nor} = 500$ Oe; and a slanted loop for Co$_{oblique}$: $M_r^{obl} = 0.78$, $M_s^{obl} = 0.5$, and $H_c^{obl} = 1600$ Oe. The deconvoluted contributions of the two magnetic components, together with the combined simulated loop fitted on the actual loop, are shown in fig. 9.

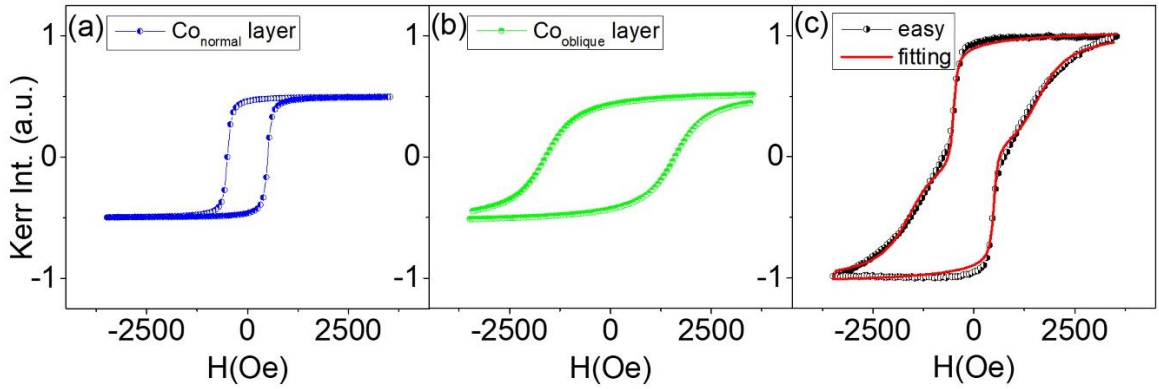

**Figure 9:** Simulated (a) Co$_{normal}$ and (b) Co$_{oblique}$ loops' contributions along the easy axis for pristine sample together with (c) combined simulated loop superimposed (fitted) on easy axis loop for pristine sample.

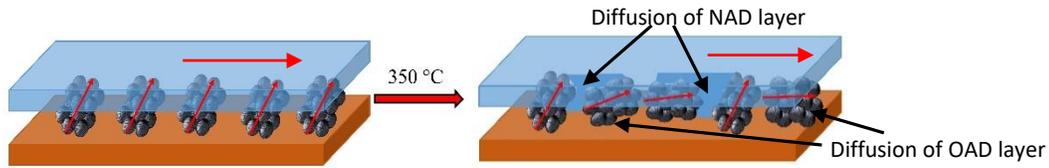

**Figure 10:** Simplified schematic of the alignment of magnetic moments (red arrows) in oblique and normal layers for pristine and annealed multilayer (350 °C). Tilted columns are represented as a tilted hcp Co structure. For presentation simplicity, only the bilayer structure is shown.

As the sample is annealed at 350 °C, a significant portion of columns diffuse to in-plane (fig. 10) direction, as confirmed by the GISAXS image of 350 °C (fig. 7d) compared to the pristine sample (fig. 7b) and GISAXS 1D data analysis (fig. 8a) also confirms the same. Even the elongation in the direct beam spot has been changed to more vertical for 350 °C (fig. 7d) compared to the pristine sample (fig. 7b), confirming diffusion. It causes the significant in-plane transition of crystallographic texturing of the hcp c-axis along the projection of columns (fig. 5d) in the multilayer plane, which is the easy axis of UMA. This process weakens the interlayer coupling between $Co_{oblique}$ layers. Thus, it provides a good square loop along the easy axis at 350 °C (fig. 2b).

To estimate the strength of UMA, the experimental value of UMA energy, $K_U$, has been calculated using the relation [13,35]:

$$K_U = M_s H_a / 2. \tag{2}$$

Here $M_s$ is the bulk saturation value (1400 emu/cm$^3$) of Co [48], and $H_a$ is the anisotropy field, determined by the field along the hard axis directions at the value of the saturation magnetization [35,49]. The available field was not enough to saturate the magnetization of the pristine sample. The anisotropy field calculated for the sample annealed at 350 °C is $H_a$=3500 Oe. The value of $Ku \approx 2.45 \times 10^6$ erg/cm$^3$ is obtained. It is to be noted that the field is not sufficient to saturate the multilayer along the hard axis at 350 °C; otherwise, the actual $H_a$ hence the actual $Ku$ is significantly higher than the calculated value. Still, the value of calculated $Ku$ is one to two orders of magnitude higher than the previous ripple structure studies on Co films [50,51,52]. For instance, Sarathlal et al. [51] have obtained $H_a \approx 120$ Oe for 18nm thick Co film, Bukharia et al. [53] have obtained $H_a \approx 100$-150 Oe for Co with thickness ranging from 5-60nm, Liedke et al. [50] have obtained anisotropy field up to 500 Oe for Co film deposited over the rippled substrate. Arranz et al. [48] have achieved the value of $H_a \approx 800$ Oe by directly modifying the surface morphology of Co thin film, having a large ripple amplitude up to 20 nm as the leading variable. While our research group created $H_a \approx 1100$ Oe using a sequential deposition erosion process for 20 nm Co film [13]. In the case of external/residual stress-induced magnetic anisotropy in Co thin film, $H_a$ not more than 40 Oe is observed [5,54]. Compared to these studies, in our case, $H_a \approx 3500$ Oe has achieved at least one order higher UMA energy $Ku \approx 2.45 \times 10^6$ erg/cm$^3$.

In the area of OAD studies, in obliquely sputtered Co films on a polymer substrate, uniaxial anisotropy $Ku = 5.6 \times 10^5$ erg/cm$^3$ for Co deposited 70° OAD, and $Ku = 2.7 \times 10^4$ erg/cm$^3$ for Co deposited 50° OAD [7] are observed. Ahmad et al. reported the value of $Ku = 2.74 \times 10^5$ erg/cm$^3$ for Co deposited at 70° [55]. C. Rizal et al. [35] reported that Co/Au multilayer deposited at various OAD angles shows the highest UMA of $2.98 \times 10^5$ erg/cm$^3$ for 45° OAD and $5.23 \times 10^5$ erg/cm$^3$ for magnetically annealed 45° OAD sample. Similarly, other OAD studies on Co also show $Ku$ of the order of $10^5$ erg/cm$^3$ [56,57,58,59,60]. While in our case, $Ku = 2.19 \times 10^6$ erg/cm$^3$ is one higher than all these studies. It suggests that the present oblique-normal sequential deposition technique imprints a significantly increased UMA.

The coercivity continuously reduces up to 450 °C and increases at 500 °C (fig. 2e) when the fcc structure dominates on the hcp Co structure [61]. This behaviour of coercivity with temperature is the result of two competing processes: (i) $H_C$ reduces with reduced pinning centres (due to reduced porosity) with increasing temperature [62,63,64], (ii) $H_C$ increases with increasing grain size/particle size [65]. GISAXS data (fig. 7 and 8) shows significant diffusion up to 450 °C, indicating that pinning centres reduce significantly as the temperature reaches 450 °C; hence $H_C$ reduces. But GIXRD analysis and AFM data (fig. 6b and 6c) confirm that as the sample is annealed to 500 °C, crystallite/grain size increases significantly, increasing the $H_C$ at 500 °C. Also, the film becomes significantly rough, as confirmed by AFM analysis and XRR data of 500 °C (fig. 6a), increasing pinning centres. Hence at 500 °C, the increased crystallite/grain size and roughness cause the increase in $H_C$. Finally, at 500 °C, the tilted columnar structure is almost diffused, as confirmed by fig. 7(f). AFM measurements also confirm the diffusion as the AFM image for the pristine sample (fig. 6b) shows grain elongation along the deposition direction, and when annealed to 500 °C (fig. 6c), bigger grain size without any preferred elongation confirms the diffusion of columns. The removal of the Bragg peak in XRR (fig. 6a) also ensures the same. Also, at 500 °C, texturing is significantly reduced, as clear by GIXRD (fig. 5f) and RHEED (fig. 3e and 3f), which has two reasons: (i) fcc phase grows at 400 °C and starts dominating the hcp phase at 500 °C. Both phases coexist at 500 °C, and (ii) columns' diffusion (fig. 7f and 8a) results in the realignment of the hcp structure (texturing). All of this causes a significant reduction of UMA at 500 °C. It is to be noted that, in the case of fcc Co, [111] direction is the easy axis of magnetization [66]. At 400 °C, fcc (111) and hcp (002) peaks nearly overlap (fig. 5g). At this temperature, whether one or both peaks are textured, the texturing is too weak (as the ring is continuous), as clear from fig. 5(d), similar to the weak texturing of the pristine sample (fig. 4b or 5b). It causes reduced contribution of MCA to the UMA as for the pristine sample.

We know that in the obliquely deposited layer, magnetic anisotropy contains out-of-plane anisotropy contribution due to tilted columns with respect to the substrate plane, which means UMA is not perfect in-plane [7,18,23] though the thickness of OAD Co is kept as low as 16.0 nm [23]. In many OAD Co studies, an out-of-plane contribution in magnetic anisotropy comes due to texturing of hcp (002) or fcc (111) peaks, which keep the easy axis away from the in-plane direction, and these studies incorporate relatively higher thicknesses (from tens to hundreds of nm) for OAD [21,22]. In the present case, we have kept the low individual thickness of obliquely deposited Co layers (4.4 nm) between the normally deposited Co layers in the multilayer stack and found the in-plane UMA [17]. Furthermore, the advantage of the combined use of the oblique-normal deposition is that a cobalt film shows the UMA when deposited normally on an oblique cobalt underlayer [19]. Moreover, not only in the OAD Co underlayer but also when thin cobalt film is deposited at normal incidence on obliquely deposited non-magnetic Ta [58] and Pt [59] underlayers, UMA is observed. In all these cases, UMA in the normally deposited magnetic layer is associated with elongated corrugations on the underlayer. Thus, in the present scenario, each OAD and NAD Co layers contribute to the resulting in-plane UMA. More importantly, the magnetic anisotropy in the present Co multilayer is one order higher than other Co-related studies in the literature. Also, high UMA is maintained up to 450 °C. Thus, the present work demonstrates the structural engineering of sequential oblique-normal depositions to induce significantly increased in-plane UMA in the Co multilayer structure.

## CONCLUSION

A unique magnetic multilayered columnar nanostructure is engineered using Co by designing successive oblique-normal deposition. The UMA obtained is one order higher than other Co-based OAD, ripple and stressed thin film/multilayer studies in the literature. This in-plane UMA with an easy axis of magnetization along the projection of the tilted nano-columns in the multilayer plane resulted from strong shape anisotropy and weak MCA confirmed by various *in situ* and *ex situ* measurements. The absence of stress nullifies any role of magneto-elastic coupling energy to the total UMA observed. The multilayer maintains magnetic anisotropy up to 450 °C, confirming good thermal stability. Significant reduction in the in-plane UMA after annealing at 500 ºC is attributed to the diffusion and merging of columns and growth of the fcc phase together with significant removal of crystallographic texturing after the heat treatment. Thus, such OAD/NAD magnetic multilayers deposited at various oblique angles and using ultra-thin layer thickness can produce significantly increased, while controllable simultaneously, in-plane UMA that can be maintained up to very high temperatures. This unique dynamic deposition study can change the quality and strength of future magnetic devices.

## ACKNOWLEDGEMENT

The authors acknowledge Mr. Anil Gome and Dr. V. R. Reddy for XRR measurements and Dr. V. Ganeshan and Mr. M. Gangrade for AFM measurements at UGC_DAE CSR, Indore. Parts of this research were carried out at the light source PETRA III at DESY, a member of the Helmholtz Association (HGF). Finally, the authors gratefully acknowledge the Department of Science and Technology (Government of India) for providing financial support within the framework of the India@DESY collaboration.